\documentclass[final,5p,times,twocolumn]{elsarticle}

 \usepackage{graphicx,psfrag}

\usepackage{amssymb}

\journal{Nuclear Instruments and Methods in Research A}

\begin{document}

\begin{frontmatter}



\title{The Jefferson Lab Frozen Spin Target}


\author[JLab]{C.D. Keith\corref{cor1}}
\ead{ckeith@jlab.org}

\author[JLab]{J. Brock}
\author[JLab]{C. Carlin}
\author[JLab]{S.A. Comer}
\author[JLab]{D. Kashy}
\author[Edinburgh]{J. McAndrew}
\author[JLab]{D.G. Meekins}
\author[JLab]{E. Pasyuk}
\author[JLab]{J.J. Pierce} 
\author[JLab]{M.L. Seely\fnref{Meyer}}
\fntext[Meyer]{Present address: Meyer Tool and Manufacturing, Inc. Oak Lawn, Illinois 60453 USA}
\cortext[cor1]{Corresponding author}

\address[JLab]{Thomas Jefferson National Accelerator Facility, Newport News, VA 23606, USA}
\address[Edinburgh]{School of Physics, University of Edinburgh, Edinburgh, United Kingdom}

\begin{abstract}
A frozen spin polarized target, constructed at Jefferson Lab for use inside a large acceptance spectrometer, is described.
The target has been utilized for photoproduction measurements with polarized tagged photons of both longitudinal and circular polarization.  Protons in TEMPO-doped butanol were dynamically polarized to approximately 90\%
outside the spectrometer at 5~T and 200--300~mK.  Photoproduction data were acquired with the target inside the spectrometer at a frozen-spin temperature of approximately 30~mK with the polarization maintained by a thin, superconducting coil installed inside the target cryostat.   A 0.56~T solenoid was used for longitudinal target polarization and a 0.50~T dipole for transverse polarization.  Spin-lattice relaxation times as high as 4000 hours were observed.  We also report polarization results for deuterated propanediol doped with the trityl radical OX063.
 \end{abstract}


\end{frontmatter}


\section{Introduction}
\label{Intro}
One of the major research initiatives taking place in Hall B at Jefferson Lab is the NSTAR program, the experimental study
of baryonic resonances. Despite decades of electron, meson, and photo-production studies, a complete and
well-characterized spectrum of excited baryonic states remains missing.    The parameters of many resonances 
(such as their mass, width, and decay couplings) are not well known, while other resonances, predicted by various 
QCD models, have yet to be experimentally verified.  Experiments with both polarized beam and
polarized target are critical to disentangling this complex spectrum of broad,
overlapping resonances \cite{Pas2009}.  Of particular importance are experiments
that combine multiple combinations of beam and target polarization.
In this article we describe a frozen spin polarized target explicitly constructed for such
experiments inside a large acceptance spectrometer.  The target can provide either longitudinal
or transverse polarization, depending on the choice of magnet used to
maintain the polarization.

The centerpiece of the Hall B instrumentation package is the CEBAF Large Acceptance Spectrometer (CLAS), a
multi-gap, high-acceptance magnetic spectrometer in which the field is generated by six
superconducting coils in a toroidal configuration \cite{Mec2003}.  This coil arrangement leaves a field-free 
region in the center of the detector that is well suited for the insertion of a 
polarized target.  One such target, dynamically polarized by continuous microwave irradiation at 140~GHz,
has been previously described~\cite{Keith2003} and has been used inside CLAS with electron
beams up to 10~nA.  However, this target features a 5~T superconducting
magnet whose geometry limits
the acceptance for scattered particles to $\pm 55^\circ$ in the forward direction, 
whereas the acceptance of CLAS spans $\pm 135^\circ$.
Furthermore, the magnetic field is by necessity parallel to the beam line and can
therefore provide only longitudinal polarization.

\begin{figure*}
\begin{center}
\includegraphics[width=5in]{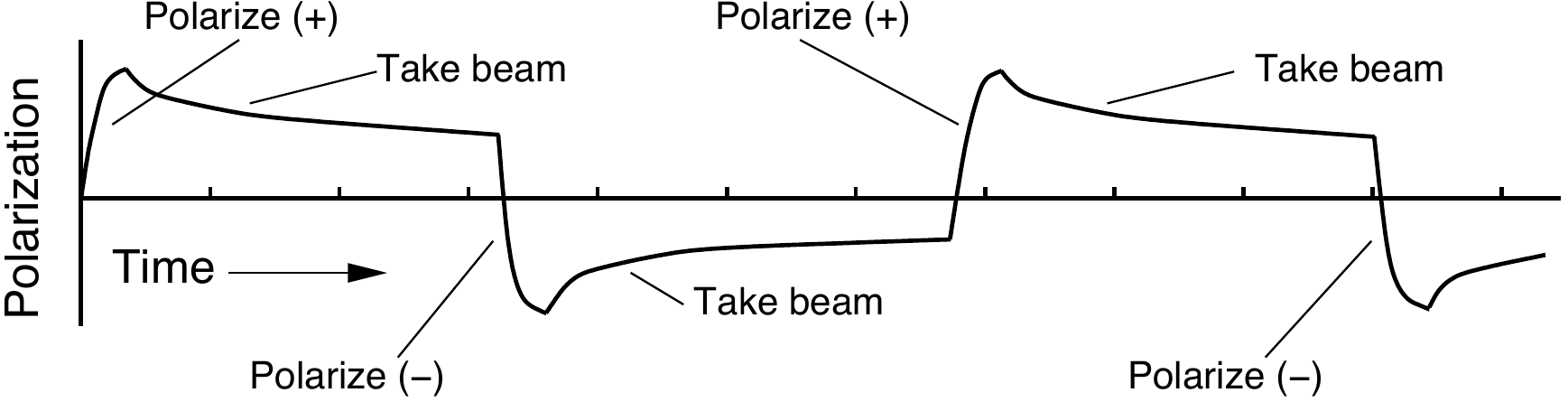}
\end{center}
\caption{Schematic depicting the operation of the frozen spin target.}
\label{FrozenCycle}
\end{figure*}

To compliment the existing polarized target in Hall B, we have designed and constructed
a frozen spin target, FROST, which permits the detection of scattered particles 
over an angular range of $\pm 135^\circ$.
The target has been utilized on two separate occasions, each lasting about six months. 
It was longitudinally polarized during the first set of experiments (g9a), and transversely polarized
for the second (g9b).   Both circularly and linearly polarized photon beams were used during g9a and g9b,
so taken together, all four possible combinations of beam and target
polarization were realized, resulting in a so-called ``complete'' experiment.

The remainder of this article is organized as follows.  A general overview of the target system
and its operation is given in Section~\ref{Overview}, with more detailed descriptions of the
various components provided in Section~\ref{Components}.  Its performance during both experiments
is described in Section~\ref{Results}, and a summary is made in the 
final section.

\section{System Overview}
\label{Overview}
The operation of the frozen spin target is depicted schematically in Fig.~\ref{FrozenCycle}.  The target sample
is polarized with microwaves via Dynamic Nuclear Polarization (DNP) in the bore of a high field, 
high homogeneity magnet.  The microwaves are then switched off,
the target is cooled to a temperature below approximately 50~mK, and the polarization
is maintained by a weaker magnetic field.  In our case this field is produced by a small superconducting
magnet installed inside the target cryostat.  
The scattering data is acquired while the target polarization decays
in an exponential manner with a spin-lattice time constant $T_1$ 
that is determined in part by the temperature of the material and the 
strength of the magnetic field.  The DNP process is repeated when
the polarization has fallen to an unacceptably low
value or to change the direction of the polarization.

The major components of FROST, which we describe in the following section, are:
\begin{enumerate}
	\item a 5~T polarizing magnet with a horizontal warm bore;
	\item a bespoke, horizontal $^3$He-$^4$He dilution refrigerator (DR) with a high cooling capacity;
	\item a novel vacuum seal and insertion device for loading the target material directly into the mixing chamber of the DR;
	\item one of two internal superconducting holding coils;
	\item a 140~GHz microwave system for the DNP process;
	\item a continuous-wave, nuclear magnetic resonance (CW-NMR) system for measuring the polarization.
\end{enumerate} 
 
\begin{figure*}
\begin{center}
\includegraphics[width=5in]{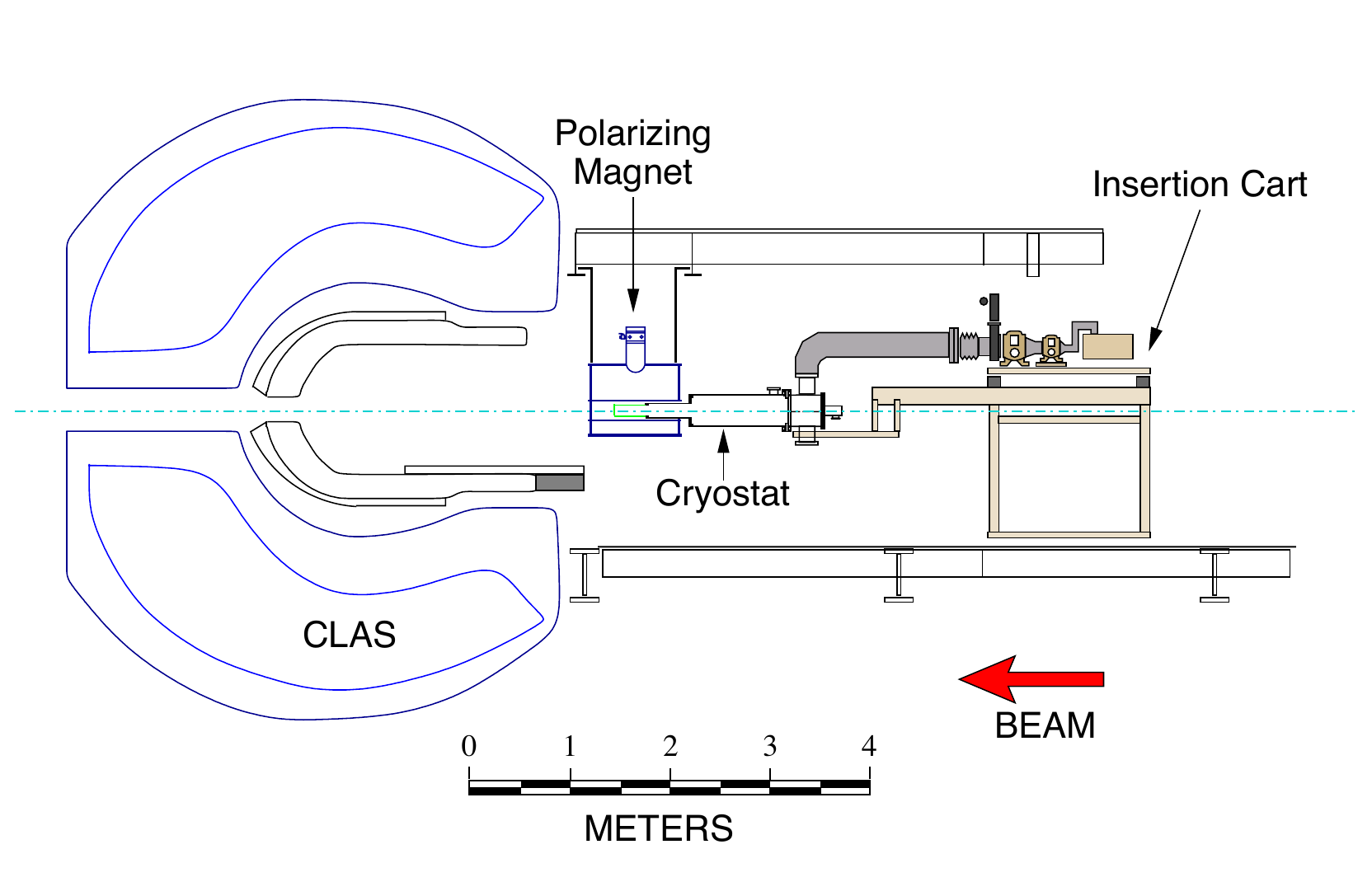}
\end{center}
\caption{Side view of the frozen spin target in the polarizing position.  To acquire photoproduction
data the polarizing magnet is moved away, and the cryostat is rolled into the center of CLAS.}
\label{FROSTside}
\end{figure*}

The layout of the system is shown in Fig.~\ref{FROSTside}.
The target cryostat is suspended from the lower portion of a two-tiered insertion cart that is
mounted on rails for travel into the center of the spectrometer.  
Most of the vacuum pumps for the cryostat are mounted to the upper tier, which is
vibrationally isolated from the lower tier by four air springs.  
A control panel for the pumps, electronics for monitoring and
controlling the target, and two gas panels for the DR are also mounted to the lower tier.  
Only a large chiller for water-cooled vacuum pumps and two tanks for storing the 
$^3$He-$^4$He mash are not mounted to the insertion cart.  These items are located on
Level 2 of the steel frame surrounding the spectrometer, and connect via flexible lines to the
target on Level 1 below.  The polarizing magnet is suspended at the entrance to the spectrometer
from a set of rails perpendicular to the beamline, and can be moved approximately 
1~m in the beam-left direction, allowing the target cart to be moved to the center of 
CLAS.  

The polarization process begins with the target cryostat inserted into the bore of the polarizing magnet (at 5~T)
and the microwave tube energized.  The time required to reach 80\% proton polarization is about 2--3 hours, 
and another 3 hours are necessary to reach 90--95\%.  During this time the microwaves
(50--100~mW) warm the target sample to about 0.3 K. The field of the polarizing magnet is parallel to the beamline, 
and so the target is longitudinally polarized. 
After the microwaves are switched off, 30--45 minutes are required for the 
target to cool to a temperature less than 50 mK.  At that time the polarizing magnet is de-energized,
while the internal holding magnet is simultaneously energized at a rate that maintains
a net magnetic field of about 0.5~T (another 45 minutes).
The direction of the polarization either remains longitudinal, or is rotated transverse
to the beam, depending on which holding coil is installed inside the target cryostat.

The cryostat is then retracted from the polarizing magnet,
the latter is moved out of the way, and
the cryostat is moved approximately 4 meters into the
center of CLAS (about 2 minutes) where it continues to cool below 30~mK. 
The tagged photon beam is activated and photoproduction data is
acquired for a period of 5--10 days, after which the polarization process is repeated, usually to
reverse the target polarization. The photon beam deposits 10--20~$\mu$W to the refrigerator, 
warming it 2 mK or so. Even with beam on target, the polarization loss is only about 1\% per day. 

\section{Components of FROST}
\label{Components}
\subsection{Polarizing Magnet}
\label{PolarizingMagnet}
A 5~Tesla solenoid\footnote{Cryomagnetics, Inc.} was used during the DNP process to polarize the target material.  
The solenoid has a horizontal, room-temperature bore of 127~mm diameter and produces a 
maximum field of 5.1~T at 82.8~A.  The inhomogeneity of the solenoid's central field
is $\le5 \times10^{-5}$ over the volume of our sample, a
50~mm long, 15~mm diameter cylinder.  It 
features a 45~l volume for liquid helium and two vapor-cooled heat shields, obviating the 
need for liquid nitrogen.  Using a flexible transfer line, the magnet was automatically filled 
with LHe about every 4 days from a nearby 500~l dewar.  Liquid in the dewar was continuously 
replenished from Jefferson Lab's End Station Refrigerator, or ESR.

\subsection{Dilution Refrigerator}
\label{DR}
\begin{figure*}
\begin{center}
\includegraphics[width=5.25in]{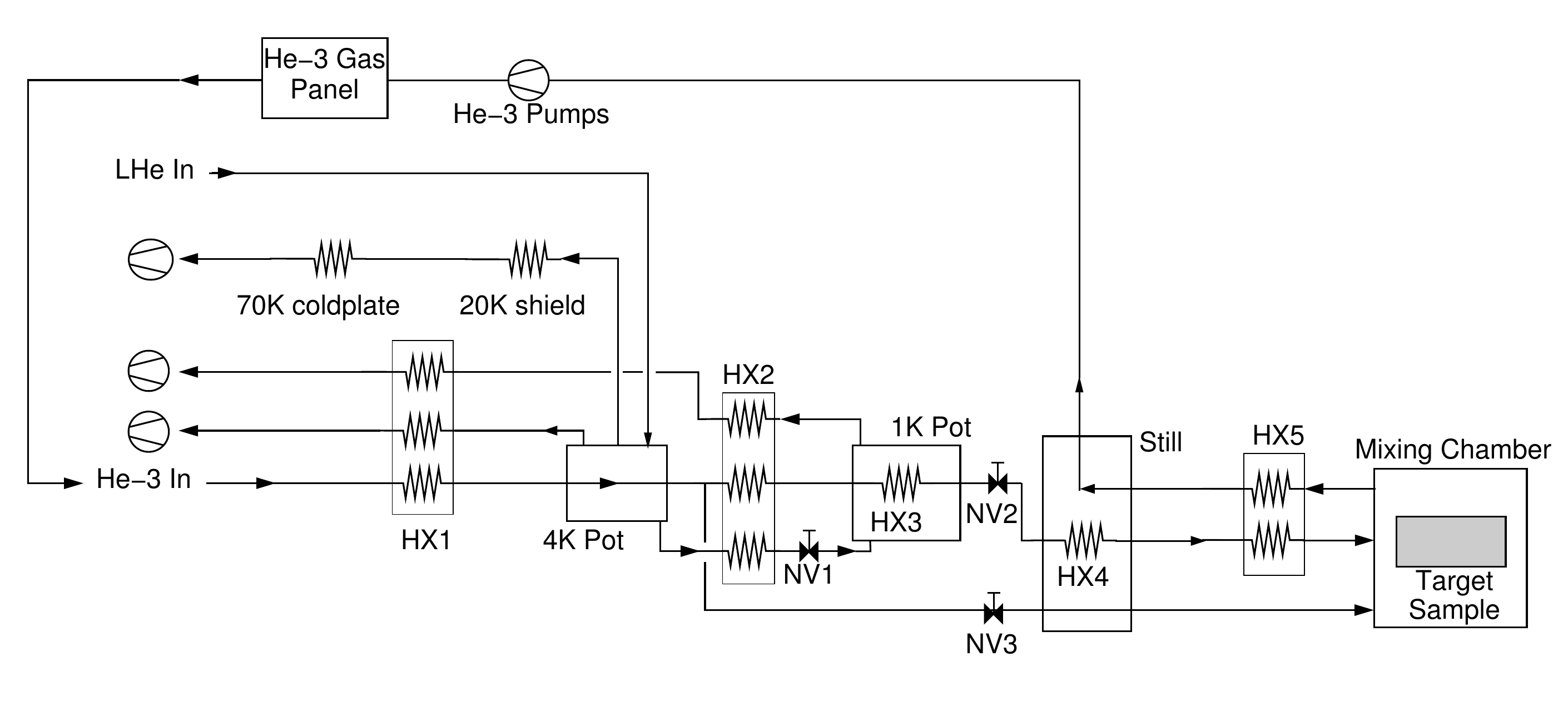}
\end{center}
\caption{Flow schematic of the target.  Here HX indicates a heat exchanger and NV a needle valve.}
\label{FlowDiagram}
\end{figure*}

\begin{figure*}
\begin{center}
\includegraphics[width=6.5in]{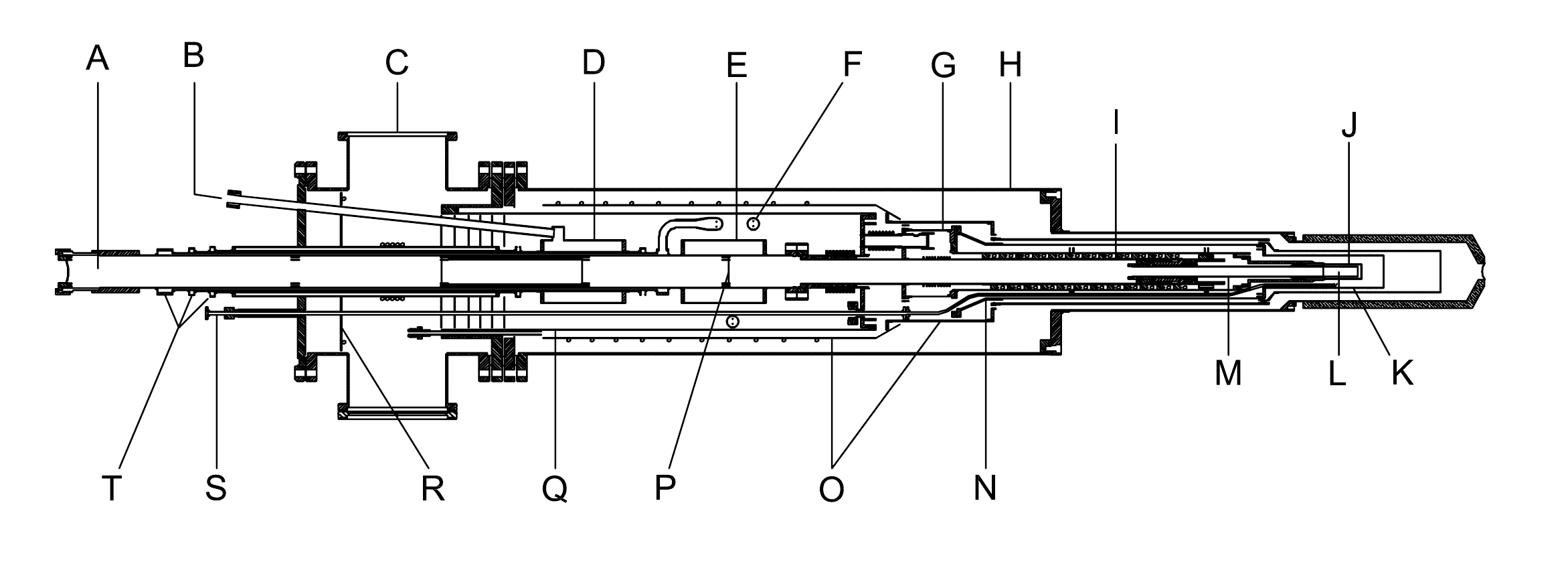}
\end{center}
\caption{Sectional view of the frozen spin target. A: beam pipe, B: LHe inlet, C: $^3$He pump port,
D: 4~K pot, E: 1~K pot, F: 1~K heat exchanger, G: still, H: vacuum chamber,
I: sintered heat exchanger, J: mixing chamber, K:  holding coil, L: target cup, M: target insert, 
N: 1~K heat shield, O: 20~K heat shield, P: beam pipe heat shield (one of three), Q: $^3$He pump tube, 
R: copper cold plate, S: waveguide, T: precool heat exchanger. 
The overall length of the cryostat is approximately 2 m.}
\label{TargetSection}
\end{figure*}

To aid in the following discussion, a flow diagram for the frozen spin target is shown in Fig.~\ref{FlowDiagram},
while a sectional view of the target is given in Fig.\ref{TargetSection}.
There are two quasi-independent refrigeration systems inside
the target cryostat: a $^3$He-$^4$He dilution refrigerator and a $^4$He evaporation
refrigerator, also called the precooler.  To conserve space, the precooler is 
located inside the 180 mm diameter tube that is used for pumping $^3$He
from the dilution unit.  It is sealed inside this pumping tube using an
indium seal at one end and a rubber o-ring at the other. Both
refrigeration units are constructed around a separate, thin-walled
stainless tube that is 50~mm in diameter in the case of the precooler, and
40~mm for the dilution refrigerator.  These bolt together using
commercial knife-edge flanges, forming a single tube that extends from
room temperature at one end to the mixing chamber at the other.  This tube
serves as both an evacuated beam pipe for the photon beam, and as a
load-lock tube for inserting the polarizable target sample into the mixing
chamber.  At the room temperature end, the tube is sealed with a 0.13~mm thick
polyimide beam-entrance window.  The opposite end is sealed by a 0.13~mm thick
aluminum window on the sample insertion device described in Section~\ref{Insert}.

As its name implies, the precooler's purpose is to cool and condense $^3$He before it
circulates through the dilution unit.  It consists of two counterflow
gas-gas heat exchangers, and two 1 liter vessels, or ``pots'',  containing
liquid helium at 4~K and 1~K, respectively.  Both pots are welded around the
central beam pipe and are instrumented with RuO temperature sensors and
miniature superconducting level probes.  The 4K pot is continuously filled
with LHe from the same 500 l dewar that services the polarizing magnet. 
Vapor is pumped from this pot along two paths.  The first is used to cool incoming $^3$He via heat
exchanger HX1, while the second cools both a 20~K heat shield surrounding the refrigerator as well
as a copper plate that is located in the upstream end of the target and
used to heat sink all tubes, wires, and cables as the enter the cryostat.
The 1K pot receives liquid helium from the 4K pot via needle valve NV1.
Vapor pumped from this pot is also used to cool $^3$He gas using heat exchangers
HX1 and HX2.  The $^3$He gas is condensed inside the 1K pot using HX3, the condenser.

HX1 consists of three, thin-walled stainless tubes welded coaxially around the
beam pipe.  The tubes are approximately 80~cm long 
with annular gaps between one tube and the next of 0.25, 1, and 2~mm.  Cold helium gas
from the 1K and 4K pots is pumped through the inner and outermost gaps,
while incoming $^3$He flows in the central gap.  In this manner the $^3$He
exchanges heat with gas from both the 4K and 1K pots and is cooled to
about 3K.  HX2 consists of a 19~mm diameter, 40~cm long 
tube for low pressure gas pumped from the 1K pot.  A pair of 3~m long, 3~mm diameter
tubes are coiled tightly inside and carry liquid helium from the 4K pot and
$^3$He from HX1.  The condenser is a 10~cm$^3$ copper cup, 
filled with sintered 50~$\mu$m copper powder and located inside the 1K pot.
The $^3$He condensation pressure is set by needle valve NV2, located downstream of the condenser.
A third needle valve, NV3, bypasses both HX2 and the condenser to deliver cold $^3$He gas
directly to the mixing chamber.  This valve is used only for initially
cooling the dilution unit from room temperature and is closed during normal operation.

Two small diaphragm pumps are used to pump vapor from the 4K pot, while a 120~m$^3$~h$^{-1}$ roots
system is used for the 1K pot.  All gas pumped from the precooler
is returned to the End Station Refrigerator for liquefaction. 

\begin{figure}
\begin{center}
\includegraphics[width=3.3in]{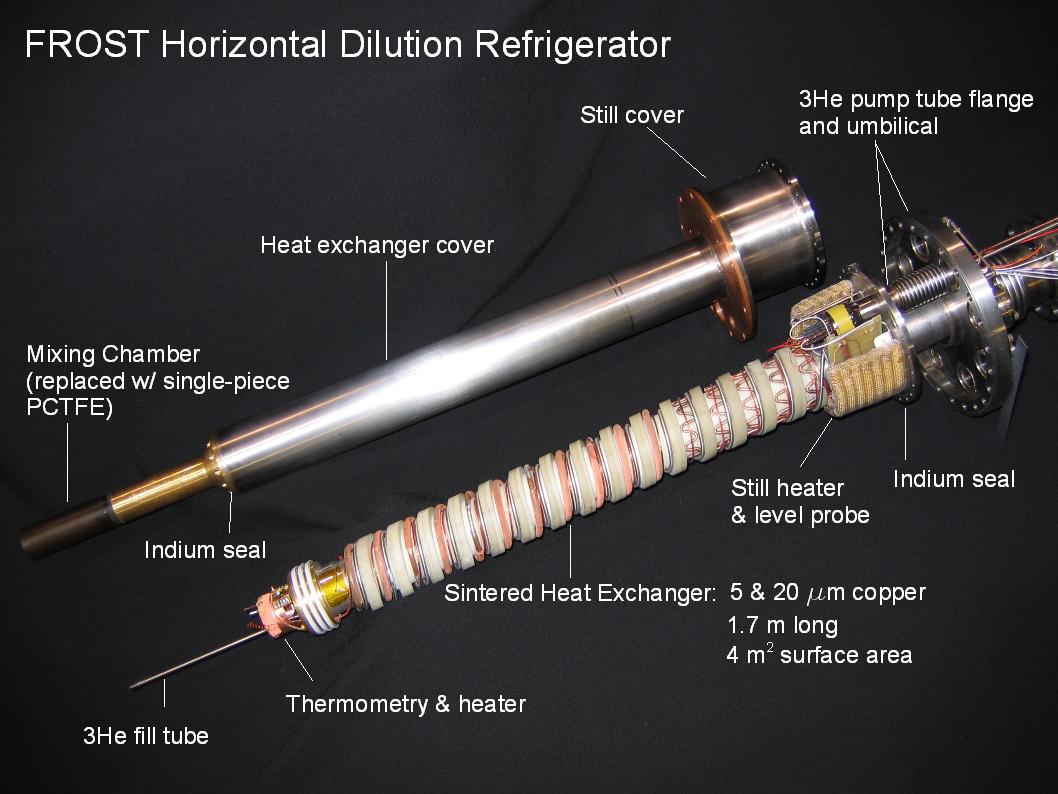}
\end{center}
\caption{Annotated photograph of the dilution refrigerator.  The brass and PCTFE mixing chamber
was subject to leaks and was replaced with one fabricated entirely from PCTFE.}
\label{DRphoto}
\end{figure}

The dilution unit consist of the customary still, heat exchanger, and mixing chamber.  It is located downstream of
the 180~mm pumping tube and is surrounded by the 20~K heat shield.  
A second heat shield, at 1~K, is attached to the still and surrounds both the heat exchanger and mixer.  
Both heat shields are constructed of copper with 1~mm thick aluminum extensions around the mixing chamber.
A photograph of the DR is shown in Fig.~\ref{DRphoto}.

The still is constructed from a 100~mm diameter stainless steel tube which seals 
with an indium o-ring against a circular flange welded around the central beam pipe.  
It is instrumented with thermometers, a heater, a capacitive liquid-level meter, and 1~m long,
1.5~mm diameter heat exchanger (HX4) for condensing any $^3$He that
may vaporize after expanding though NV2.  The heater is a 10~m long coil of 0.6~mm NiChrome wire.
The level probe consists of two copper-clad fiberglass plates with a separation of 0.5~mm, and
the capacitance between the plates is measured using an AC bridge circuit.

Gas from the still is pumped through a short, flexible bellows of 16~mm inner diameter
connecting the top of the still to the 180~mm diameter pumping flange.  
All wiring for the DR, along with the $^3$He bypass and $^3$He condenser lines, 
pass through this bellows, or ``umbilical''.  $^3$He is pumped from the still by two dry pumping systems
operating in parallel.  Each system consists of Alcatel RSV2000 
and RSV600 roots pumps and an Edwards L70 dry pump, and the measured pumping speed for
helium is 3300~m$^3$/hr.  The $^3$He gas panel features two liquid nitrogen traps for filtering contaminates
from the circulating gas (the second trap is a spare), 
but thanks to the all-dry pumping system, the trap has blocked only once, despite months of
continuous use.  The contaminates on that occasion were traced to a room-temperature gas fitting that
had presumably vibrated loose.

The main heat exchanger for the DR (HX5) is modeled after a design by Niinikoski~\cite{Niin1976} 
and is comprised of three sections.  The first is a
2~m length\footnote{Later shortened to 1~m, see Section~\ref{FridgePerformance}}
 of cupronickel tube with an inner diameter of 0.5~mm. This
serves as a secondary flow impedance for the circulating
$^3$He and was added after the refrigerator experienced flow and temperature instabilities during early tests.
The second section of the heat exchanger is a 1.5~m long stainless steel tube of 1.0~mm inner diameter
with copper fins brazed to the outside.\footnote{Fin Tube Products, Inc.}
The final section of heat exchanger is comprised of six 
copper C122 tubes with copper powder sintered inside and out.  
Each tube is 28~cm long, with inner and 
outer diameters of 3.3 and 4.0~mm.  The sinter on the outside is 1~mm thick, 
while the inside is filled with sinter except for a central, 1.5~mm diameter flow
channel for the concentrated $^3$He stream.  
The warmest two sections are sintered with 
325 mesh copper powder (nominal size 20~$\mu$m), while
5~$\mu$m powder was used on the coldest four.  
The total surface area of the sintered heat exchanger
is about 4~m$^2$ on the concentrated side and 7~m$^2$ on the dilute side, based on 
77~K measurements of argon gas adsorption.

All sections of the heat exchanger are wound in a spiral groove
machined in a G10 fiberglass mandrel.  The groove is 3.5~m long and has a
13~mm wide, 7~mm deep rectangular cross section.  All wires to the mixing chamber and the 
3~mm bypass tube are also wound in this groove.
The G10 mandrel slides tightly over the central beam pipe and is covered by
a second tightly-fitting stainless steel tube that attaches to the still at one end and to the mixing chamber at the other.
The mixing chamber is a 1~mm thick PCTFE cup that seals against this tube using
an indium o-ring.  Concentrated $^3$He enters the mixing chamber through a PTFE tube 
located in the lower half of the mixer, directly under the target sample.  A series of holes are
punctured along the length of the tube and distribute the concentrate evenly as it rises past the
sample and collects at the top of the mixer.  
A series of three spring-energized PTFE radial seals constrains the excess concentrate to
the top half of mixing chamber, while 3~mm holes under the seals allow $^3$He to be removed
from the dilute phase in the lower half.

The mixing chamber is outfitted with a small nichrome heater for cooling power measurements and
three RuO resistance thermometers.  One thermometer,\footnote{Lakeshore RX-202-AA}
has a room temperature resistance of 2~k$\Omega$ 
and is calibrated 0.05--40~K.  However, its resistance versus temperature dependence
is too steep to permit accurate extrapolation to lower temperatures.  
The additional two resistors\footnote{Dale 1 k$\Omega$  RCW-575} were chosen because they are
known to exhibit a well-behaved, log(T$^{n}$) response at temperatures as low as 25 mK \cite{Uhlig1995}
and are calibrated against the first thermometer.  
One of these is located in the center of the mixing chamber, near the polarized target sample,
while the other is in the downstream end of the heat exchanger, in the dilute flow stream.
All temperatures quotes in this article are based on this last sensor as it is believed to measure the
average temperature of liquid exiting the mixing chamber.  The thermometers in the mixing chamber and in the still
are read by a Lakeshore Model 370 AC resistance bridge, which is also used to power both the
still and mixing chamber heaters.

Both the dilution and precooling refrigerators are housed inside a custom-built,
stainless steel vacuum chamber.  The downstream end of the chamber
is made of closed-cell foam to reduce the energy loss of particles scattered from the 
polarized butanol sample.  A thin layer of aluminum is glued to the foam's interior surface
to decrease outgassing, and a 25~$\mu$m thick aluminum exit window is glued to its
downstream end.

\subsection{Target Material and Insert}
\label{Insert}
Frozen beads of butanol (C$_4$H$_9$OH) were used for the target material.  The butanol was doped with the
nitroxyl radical TEMPO\footnote{2,2,6,6-Tetramethylpiperidinyloxy} at a concentration of 
$2.0 \times 10^{19}$~spins~cm$^{-3}$
for dynamic polarization.  Water (0.5\% by weight) was added to the solution before freezing in order
to avoid a crystalline solid.  The 1--2~mm diameter beads were formed by dripping the solution
through hypodermic needles into a bath of liquid nitrogen.  The needles were held at an
electrical potential of approximately 2~kV in order to control the bead size.
This also provided the beads with a slight static charge
which kept them from clumping together in the liquid nitrogen bath.

Approximately 5~g of beads were loaded under liquid nitrogen into a 
15~mm diameter, 50~mm long PCTFE target cup that attached to end
of a 25~cm long stainless steel tube.  The tube is sealed at the target end by
a 0.13~mm thick aluminum vacuum window that also features a locking
mechanism for the PCTFE cup. 
At the opposite end, a flange for a novel vacuum seal (described below) is welded to the tube.  
The cup has a number of 0.5~mm wide grooves machined
into its underside which allows some of the $^3$He concentrate entering the mixing 
chamber to penetrate directly into the cup.  Since $^3$He absorbs heat when it dilutes into liquid $^4$He, 
we believe this may provide a more efficient cooling path to the target beads.

We have developed a new superfluid-tight vacuum seal for quickly loading the target material
into the mixing chamber of the dilution refrigerator while at cryogenic temperatures.
Patterned after VCR face-seal fittings,\footnote{Swagelok, Inc.}
it features a 0.6~mm high, toroidally-shaped ridge on the insert face for 
compressing a polyimide vacuum gasket.  A threaded nut is used to screw
the insert against the cryostat-side sealing face which has a
polished, conical sealing surface and male threads.  The conical design ensures that thermal contraction
continues to tighten the seal after the warm insert (80~K) is screwed into the cold cryostat (10~K).
A set of bearings between the insert's face seal and its threaded nut prevents the insert from rotating 
while the nut is turned.

The seal is made by attaching the target insert, under liquid nitrogen, 
to a 2~m long pipe that acts as a long wrench.  This pipe has
a set of retractable pins that lock into matching holes in the insert.  We load the insert into the empty
mixing chamber at a temperature of about 10~K with a strong helium purge on the beam pipe. 
After making the seal, the pins are retracted and the wrench is removed.
A set of three 13~$\mu$m thick aluminum heat shields is placed inside the beam pipe which is then
evacuated and circulation of the
$^3$He-$^4$He mixture through the dilution refrigerator is begun.  The entire procedure requires about 90 minutes,
during which the mixing chamber temperature rises to about 60--70~K.

 We have tested several gasket materials and have found that 180~$\mu$m thick polyimide gives the
 most reliable results.  We have made the seal several dozen times with almost 100\% success.  Leaks
 are almost always due to contaminants such as moisture
 on the gasket or threaded surfaces.  These contaminants are reduced by
 establishing a separate helium purge on the insert as soon as it is removed
 from the liquid nitrogen bath.

In addition to the polarized butanol sample, two additional targets were installed in the cryostat and were
used for background and dilution studies.  A 1.5~mm thick carbon disk and a 3.5~mm thick CH$_2$ disk
were mounted on the 1K and 20K heat shields, approximately 6~cm and 16~cm downstream of the butanol sample.

\subsection{Holding Coils}
\label{HoldingCoils}
\begin{figure}
\begin{center}
\includegraphics[width=3.3in]{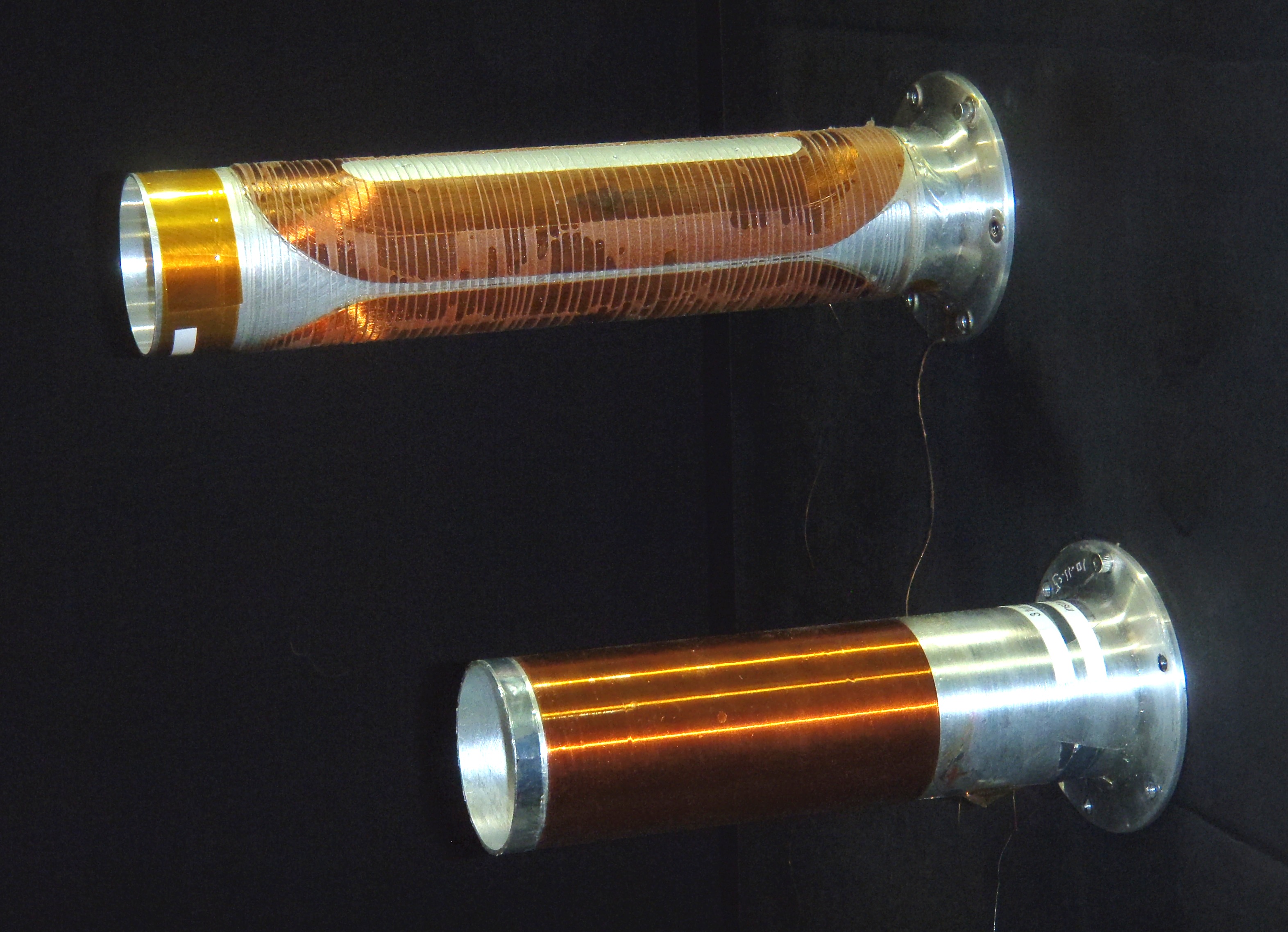}
\end{center}
\caption{Superconducting holding coils used to maintain target polarization in frozen spin mode.
Top: Four layer dipole used during g9b.  Bottom: Three layer solenoid used for g9a.}
\label{Coils}
\end{figure}
In the first generation of frozen spin targets, the holding field was generated by the fringe of the polarizing
magnet or by a second, external magnet with a larger aperture.  The field uniformity in these cases was not sufficient to
resolve the NMR line of the target material, and so it was not possible to monitor the target polarization during the
scattering experiment.  Niinikoski first described a thin superconducting solenoid mounted inside the
target cryostat for maintaining or rotating the target polarization~\cite{Niin1978}, while the group at Bonn
were the first to implement this type of solenoid in an actual experiment~\cite{Dutz1995}.  
With careful design, these coils can be thin enough for reaction products to pass through
with acceptably low energy loss and uniform enough for precise NMR measurements.

We have utilized two such coils for the g9a and g9b experiments, as shown in Fig.~\ref{Coils}. 
For g9a, a 110~mm long solenoid was used to maintain longitudinal polarization.
It consists of three layers of 0.14~mm copper-clad, multifiliment NbTi
wire\footnote{Supercon, Inc. type 54S43.} wound on a 50~mm diameter, 1~mm thick aluminum former.  
Each layer has 785 windings, while an additional 162 turns are added at both ends to improve the field uniformity.
Grooves machined in the former aid in the placement of the windings.
Stycast epoxy 1365-65N was used to adhere the coil, which produces a 0.56~T central field at 22.0~A.
The aluminum former was attached to the downstream end of the 1~K copper heat shield which was in turn
thermally sunk to the still.  

The second coil, used for g9b, was wound as a four layer, saddle-shaped dipole.  The two innermost layers each
consist of 170 turns of wire, while 152 and 60 turns were used for the third and fourth layers, respectively.
All eight coils were wound from a single, continuous length of 0.18~$\mu$m diameter superconducting wire
using a custom-designed aluminum fixture coated with PTFE.  The coils were glued to a 5~$\mu$m polyester backing to
prevent the layers from separating as they were wrapped around and epoxied to an aluminum former similar to the one
described above.  In addition, a thin nylon thread was wrapped around the coils to further secure them to the former.
This magnet has demonstrated a maximum field of 0.54~T,
and was operated at 0.50~T (35.5~A) during g9b.
The technique and equipment for winding this dipole were developed and built
at Jefferson Lab for FROST, and later used to construct similar coils for a
polarized target at the  Hi$\gamma$s facility \cite{Seo2010}.  

\begin{figure}
\begin{center}
\includegraphics[width=3in]{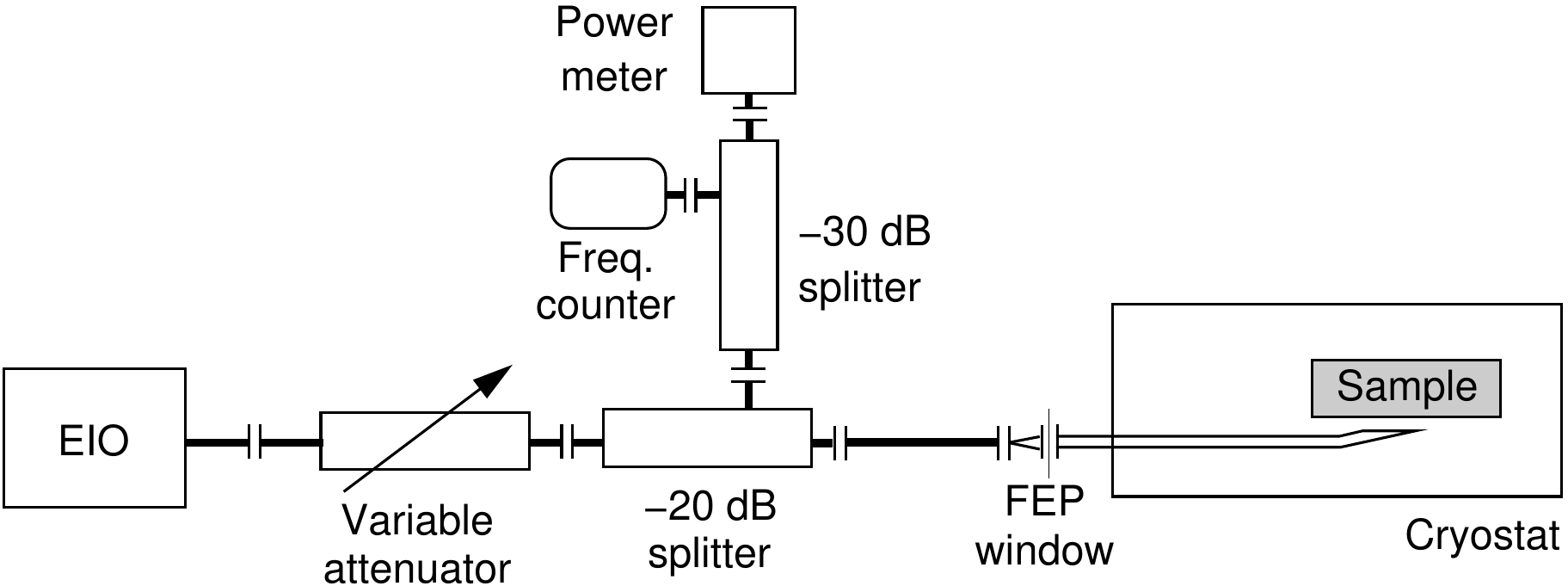}
\end{center}
\caption{Microwave system used during dynamic polarization of the frozen spin target.  
The variable attenuator was not used during g9a.  
Instead, the microwave power was adjusted with the EIO cathode voltage.}
\label{MicrowaveFig}
\end{figure}
Current leads for both coils consisted of a combination of copper wire, high temperature superconducting
ribbon,\footnote{American Superconductor HTS Cryoblock wire}  and 0.40~mm diameter NbTi wire.\footnote{Supercon T48B-M}
The leads were heat sunk at both the 4K and 1K pots, and at the still, where they were soldered to excess wire
from the holding coil.
Both the solenoid and dipole magnets quenched at full field during the g9a and g9b runs.
This happened on four occasions with the solenoid and once with the dipole.  
In the former case the quenches were eventually traced to a loose electrical
connection at the room temperature feedthrough for the magnet leads.  No quenches occurred after this was tightened,
and the solenoid was at no time damaged.
The dipole quenched when a water accident in the experimental hall shut off electrical power to the cryostat.  In this
instance one of the coil leads broke near the still heat sink.  Fortunately enough excess wire remained to remake the
solder joint at the still without having to splice in a new section of lead.

From the widths of the NMR lines obtained with the two coils (see Fig.~\ref{NMRcurves}),
we estimate that $\Delta B/B$ for the solenoid field was about 0.003 and about 0.008 for the dipole.

\subsection{Microwaves}
\label{Microwaves}
Microwaves for dynamically polarizing the target were generated by an Extended Interaction 
Oscillator\footnote{Communications \& Power Industries Canada, Inc.} with a 140~GHz center 
frequency, a tuning range of approximately $\pm$1~GHz, and a maximum power output of about 15~W.
The frequency was adjusted by changing the size of the oscillator cavity using a small DC motor. 
Microwaves were transmitted to the target sample through a 2~m length of 4.3~mm diameter
cupronickel waveguide that was sealed with a FEP window at the room temperature end
and heat sunk at several locations inside the cryostat. It terminated outside the mixing chamber
with a slight, upward bend that directed microwaves at
the target sample.  The aluminum former for the holding coil acted in some respect as a multimode cavity
for the microwaves, but its efficiency is unknown and likely quite poor.  Because of this, the best polarization
was found to result from a substantial microwave power of about 50--100~mW to the target.
A schematic diagram for the microwave system is shown in Fig.~\ref{MicrowaveFig}.

\subsection{NMR System}
\label{NMR} 
\begin{figure}
\begin{center}
\includegraphics[width=3in]{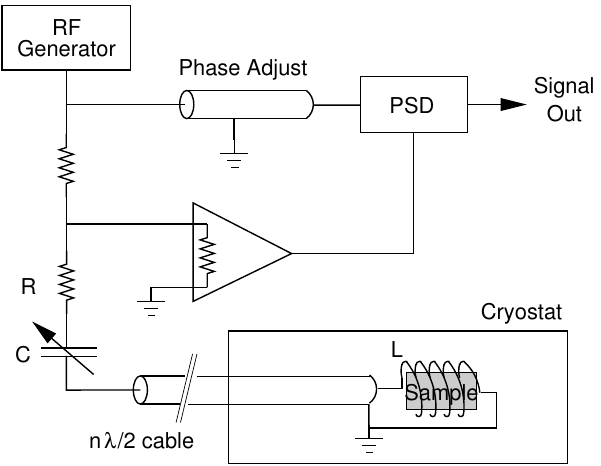}
\end{center}
\caption{Schematic diagram of the Q-meter circuit used to measure the target polarization.}
\label{NMRcircuit}
\end{figure}
The target polarization was measured using continuous-wave NMR circuits designed around the Liverpool Q-meter
\cite{Court1993}.  As shown in Fig.\ref{NMRcircuit}, this consists of a series-tuned LRC circuit
where the inductance takes the form of a small coil surrounding the target sample
while a variable capacitor is used to adjust the resonance frequency.
The RF field is swept at constant current through the nuclear resonance frequency, and the net 
energy absorbed or emitted by the target spins is observed as a change to the coil's 
impedance.  Phase-sensitive detection (PSD) is used to measure the real part of the voltage across the circuit during the
RF sweep.  The polarization of the sample is proportional to the area under the resonance peak, with
the constant of proportionality determined by calibrating the Q-meter system against a known sample polarization.

Two NMR systems were utilized during the photoproduction measurements.
The first system was tuned to resonate at 212.2~MHz and measured the target polarization at 5.0~T during the DNP
process.  This system was calibrated against the thermal equilibrium polarization of the target sample, which was measured at
various temperatures between 0.9 and 1.8~K.  The second NMR system was used to monitor the polarization
of the target while in frozen spin mode.  During g9a it was tuned at 23.8~MHz
for use with the 0.56~T solenoid, and at 21.9~MHz for the 0.50~T dipole used during g9b.
This second system was calibrated against the 212~MHz measurements each time the target was
polarized.  Comparison of the two systems is discussed in Section~\ref{PolResults}.
\begin{figure*}
\begin{center}$
\begin{array}{cc}
\includegraphics[width=2.in]{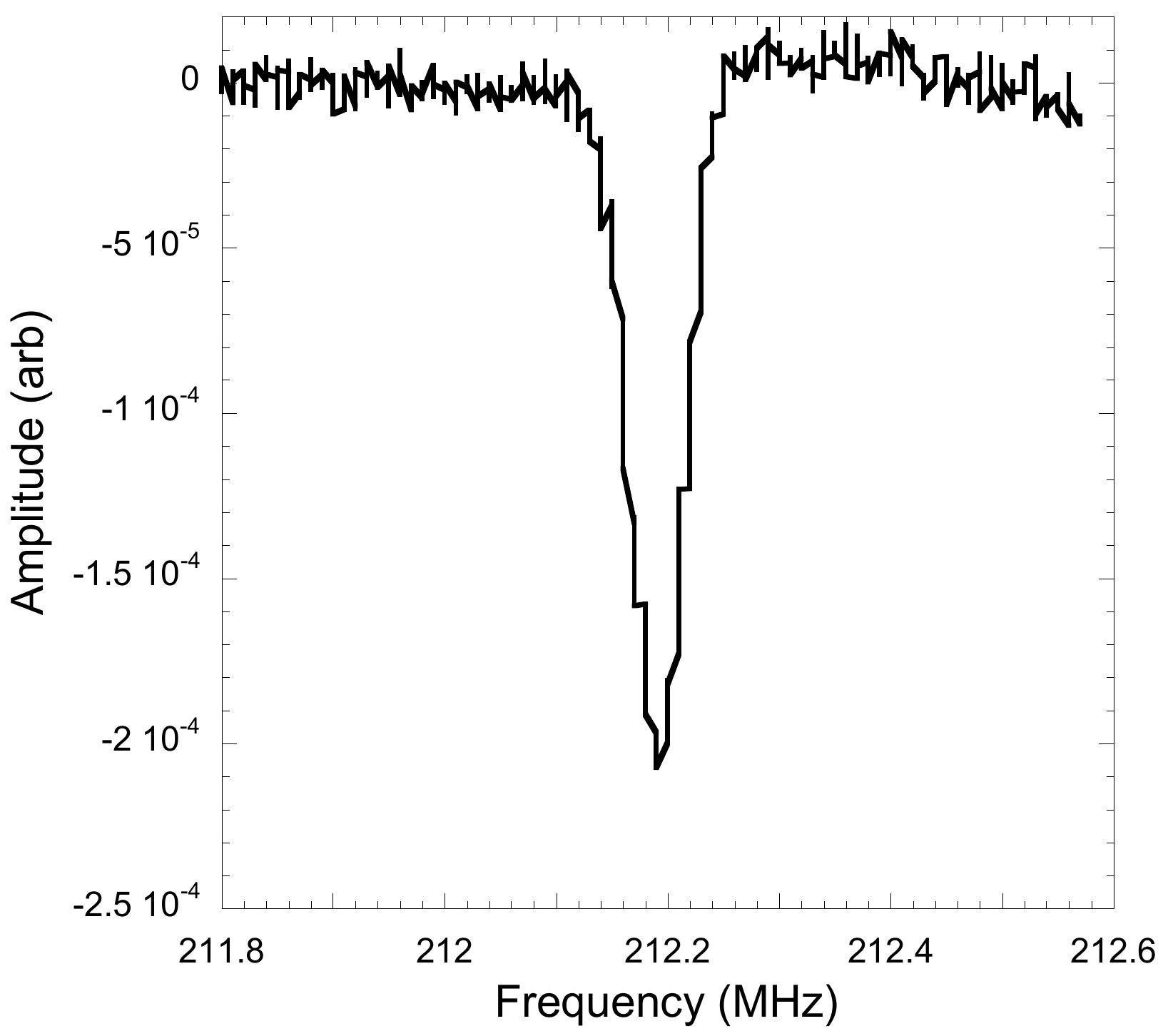} &
\includegraphics[width=2.in]{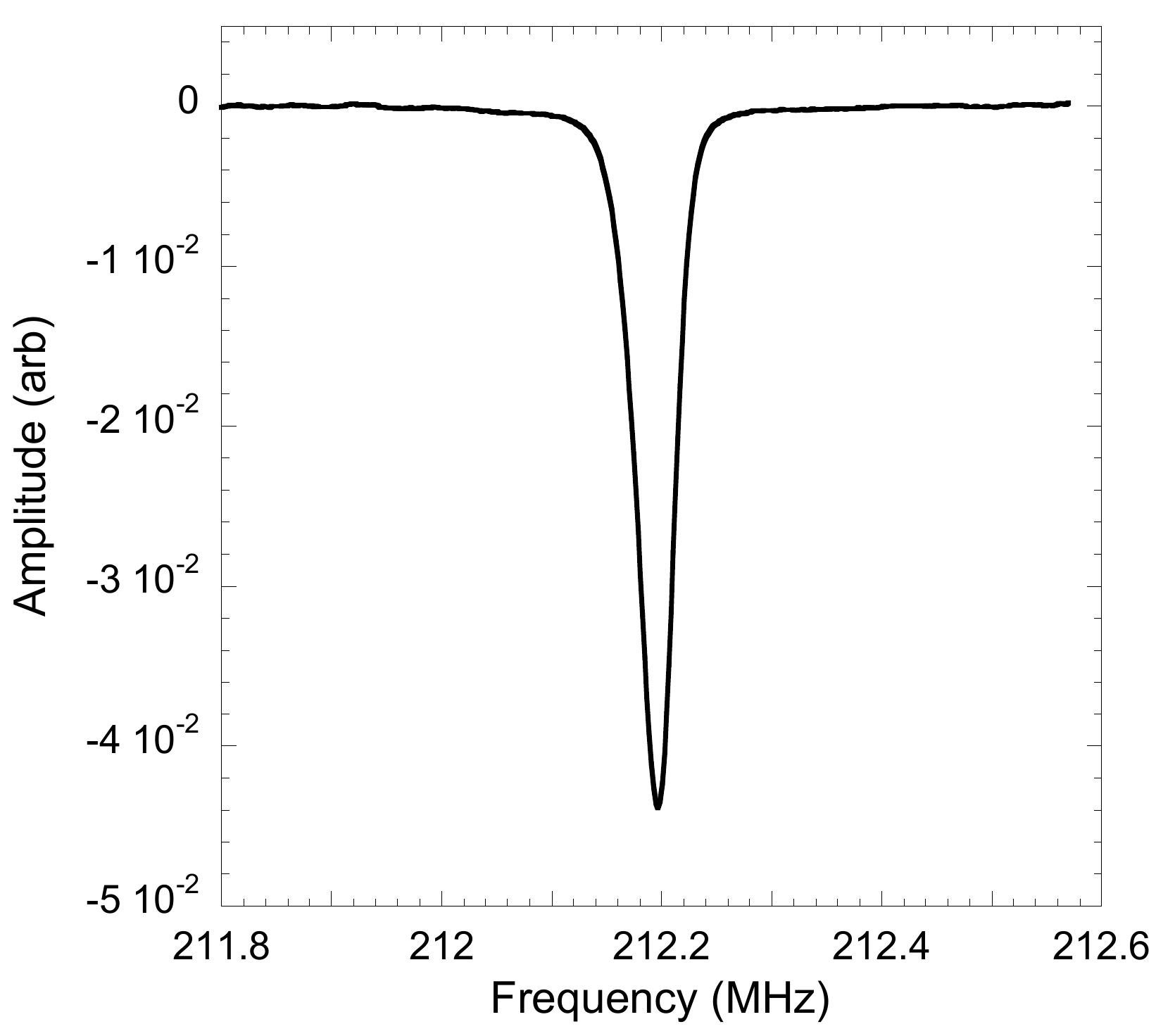} \\ 
\includegraphics[width=2.in]{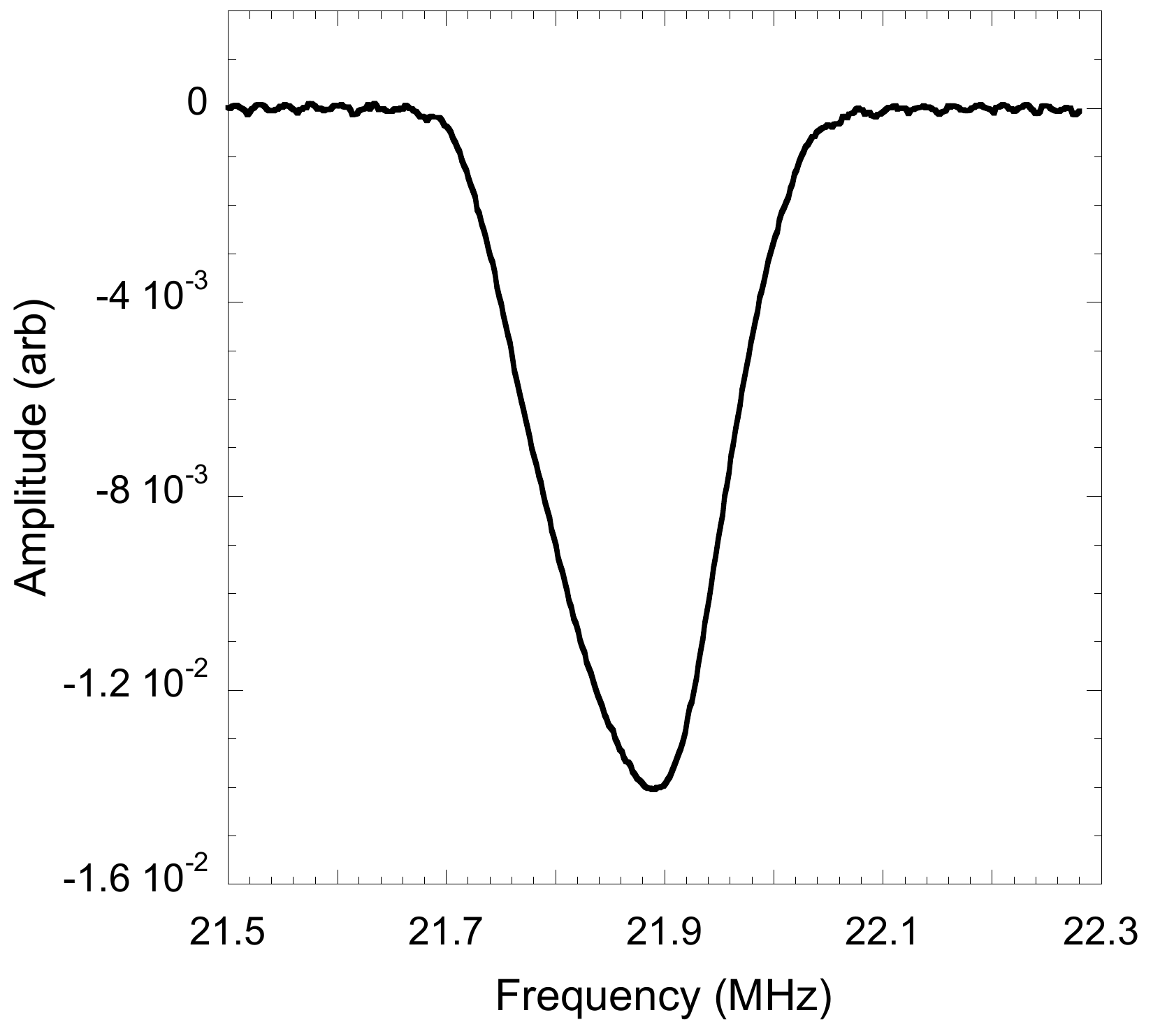} &
\includegraphics[width=2.in]{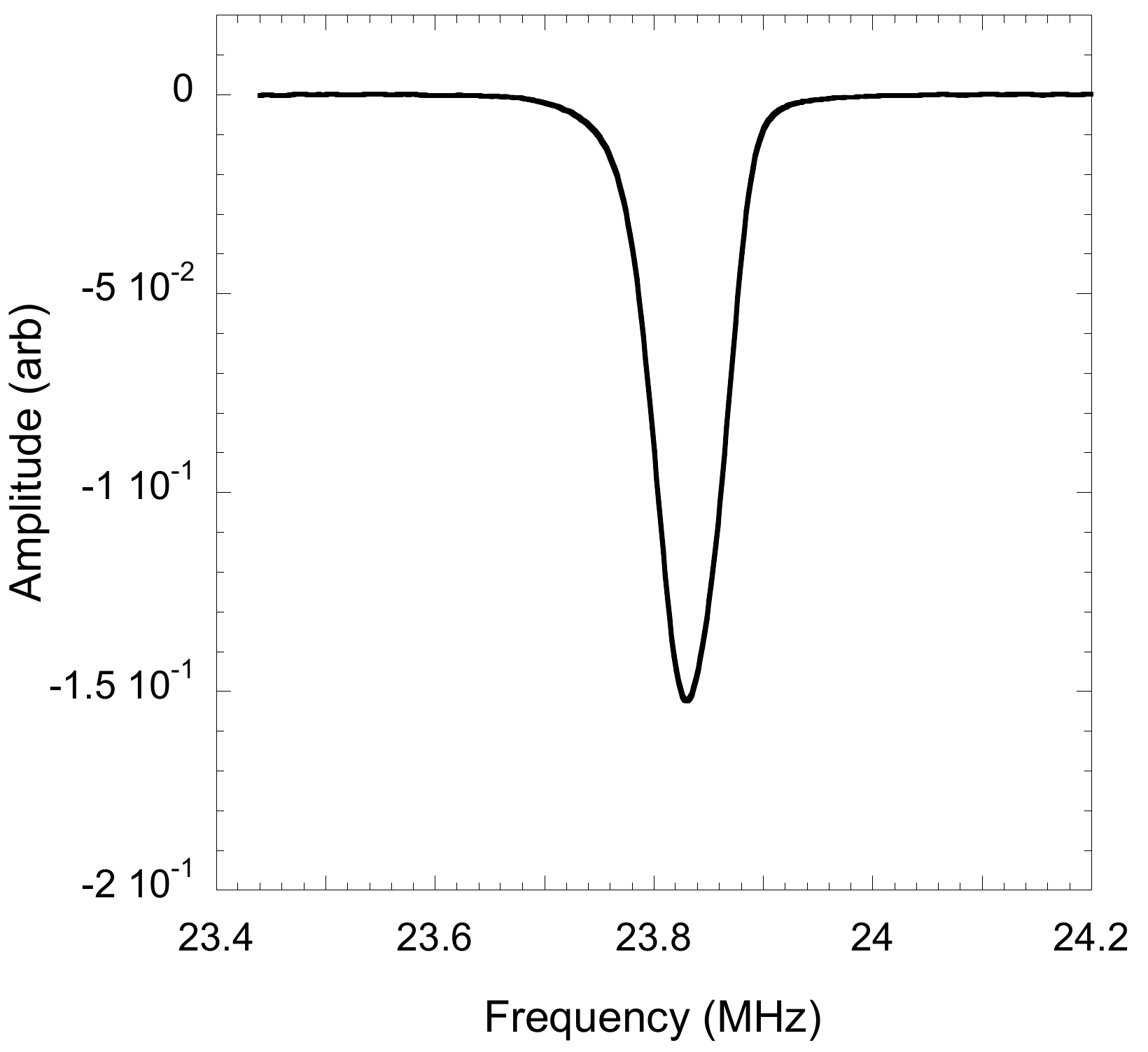}
\end{array}$
\end{center}
\caption{NMR scans of the butanol sample.  Clockwise from upper left: thermal equilibrium polarization
at 1~K and 5~T; approximately 85\% polarization at 5.0~T; longitudinal holding coil at 0.56~T;
transverse holding coil at 0.50~T.}
\label{NMRcurves}
\end{figure*}

For the thermal equilibrium calibration measurements, the DR was operated as a simple $^4$He evaporation
refrigerator with a base temperature of approximately 0.75~K.  This ensured a uniform sample temperature 
since the mixing chamber was filled with superfluid helium.  
The 2~k$\Omega$ RuO sensor in the mixer was used to determine the sample temperature with an accuracy of
2\%.  A small ($\sim$4\%) correction was applied to account for the sensor's magnetoresistance at 5~T, previously
measured against the $^3$He vapor pressure curve.

The NMR coils were cut from 25$\mu$m thick copper foil and were wrapped around the outside of the mixing chamber.
They were held rigidly in place by a thin layer of FEP heat shrink tubing.  Two separate coils were used for g9a, while
a single coil was utilized for g9b.  This is discussed in further detail in Section~\ref{Results}.  The NMR coils were
connected to the Q-meter circuit using resonant lengths ($n\lambda / 2$) of semi-rigid coaxial cables
specified for use at cryogenic temperatures.  To first order, the $n\lambda / 2$ cables 
mirror the coil inductance directly to the rest of the Q-meter circuit.  However, they also generate
a large background signal (Q-curve) which must be removed for precise signal analysis.  This is accomplished
by performing NMR sweeps with the magnetic field shifted slightly off resonance and subtracting the
resultant Q-curve from subsequent, on-resonance scans.

A typical NMR measurement consisted of sweeping through the resonance line
multiple times and averaging the results in order to improve the signal-to-noise ratio.  
The sweep width was typically $\pm$400~kHz for both the high and low field systems.  
The polarization was measured continuously during the DNP process, and twice per hour while in frozen
spin mode.  Typical NMR curves are shown in Fig.~\ref{NMRcurves}.

\section{Results}
\label{Results}
\subsection{Refrigerator Performance}
\label{FridgePerformance}
\begin{figure}
\includegraphics[width=3in]{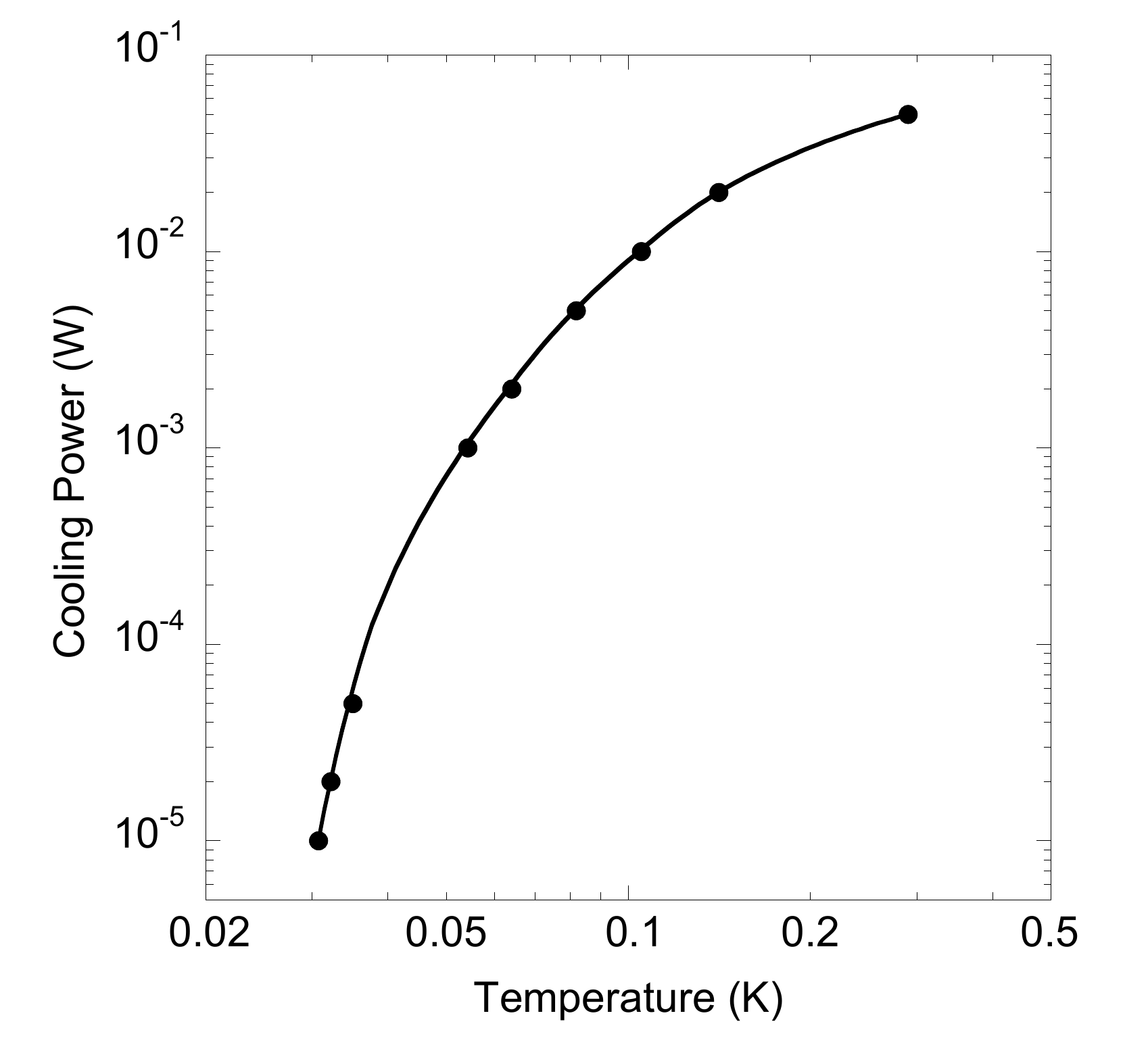}
\caption{Measured cooling power of the dilution refrigerator with $^3$He flow rates 1--16~mmol~s$^{-1}$.  
The solid line is a guide to the eye.}
\label{CoolingPower}
\end{figure}

When mounted on a concrete floor during its initial tests, 
the dilution refrigerator achieved a base temperature of 26~mK.
Vibration was more problematic in the experimental hall, resulting in a base temperature of 28~mK 
(during g9a) with a $^3$He circulation rate of 1.0~mmol~s$^{-1}$.  The maximum sustainable
flow rate during g9a was about 16~mmol~s$^{-1}$ and was
limited by the circulation pressure.
Exposed to a photon flux of about $5 \times10^{7}$~cm$^{-2}\cdot$s$^{-1}$, the target warmed to approximately 30--32~mK
at an optimal flow rate of 1.6~mmol~s$^{-1}$.
The cooling power of the refrigerator was measured during g9a with flow rates up to 16~mmol~s$^{-1}$
and is shown in Fig.~\ref{CoolingPower}.  The cooling power was approximately 1~mW at 50~mK,
10~mW at 100~mK, and 60~mW at 300~mK.

Two modifications were made to the DR between g9a and g9b.  First, the secondary flow impedance was
shortened from 2~m to 1~m, permitting the refrigerator to run with flow rates in excess of 30~mmol~s$^{-1}$.
Second, a length of PTFE cord was placed in the G10 spiral, alongside the
heat exchanger.  This reduced the axial conduction of heat through the dilute solution and lowered the 
base temperature without beam to 24--25~mK and about 28~mK with beam.
The cooling power was not measured following these modifications, but we estimate that the
increased flow rate improved the cooling power to about 100~mW at 300~mK.

As mentioned above,  a 4~m long, flexible transfer line was used to supply liquid helium to the target cryostat
from a 500~l dewar.  During g9a the efficiency of this line was very poor, and the liquid helium consumption of
the target was about 20~l~h$^{-1}$.  A new transfer line with better thermal insulation was constructed for g9b,
and the consumption dropped to  8--10~l~h$^{-1}$.

The refrigerator operated continuously and without incident for the entire length of the g9a experiment, about 5 months.
It ran six months for g9b, however its performance was compromised by
a water accident in the experimental hall during the final weeks.  This accident shut off power to the cryostat pumping system, 
damaging the transverse holding coil as noted above.
The coil was repaired but a superfluid leak between the mixing chamber and beamline appeared on the
subsequent cool down.  Attempts to repair the leak in a timely fashion were unsuccessful, and the
final two weeks of g9b were run with a base temperature of only 60~mK. 

\subsection{Polarization Results}
\label{PolResults}
The target was polarized a total of 21 times during g9a, with an average starting polarization of 84\% in the positive
spin state (9 times) and -86\% in the negative.  
Typical relaxation times for positive polarization during g9a 
were about 2800 hours with beam on target and up to 3600 hours without beam.
The target relaxed more quickly in the negative spin state, about 1400 hours with beam and 1900 hours without.
The maximum polarization was -94\%.  The target was re-polarized (and the polarization reversed)
about once per week.  

During g9b the target was polarized a total of 19 times, again about once per week.  
The target was most often polarized in the negative spin state (15 times) because it
reached a higher starting polarization (-92\%) than the positive spin state (83\%).  
Therefore the orientation of the target spins with respect to the beam was
usually determined by the direction of the transverse holding field.
The relaxation time during g9b was somewhat higher, about 3400 hours for positive polarization
with beam and 4000 hours without.  The relaxation time for the negative spin state was once again
about half that of the positive.  
The final two weeks of g9b were run with a superfluid leak between the mixing chamber and
beam pipe.  This reduced the starting polarization (three polarization cycles) to an average of only 69\%,
and the relaxation time decreased by about a factor of seven, necessitating bi-weekly polarizations.

\begin{figure}
\includegraphics[width=3in]{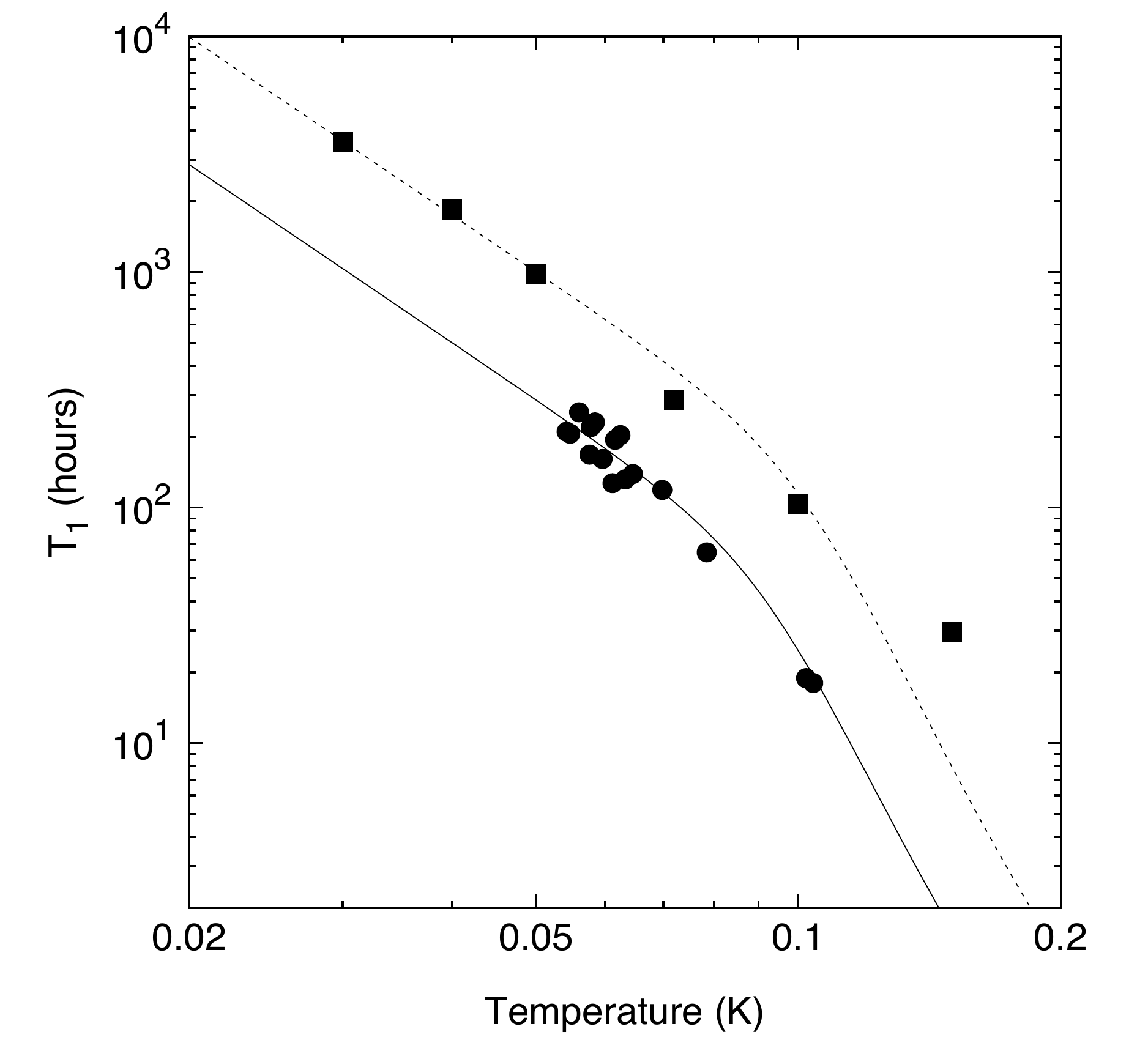}
\caption{Spin-lattice relaxation times $T_1$ of polarized butanol as a function of sample temperature for two values of magnetic field.  Circles:  Bonn frozen spin target at 0.42~T.  Squares: JLab frozen spin target at 0.56~T.  The solid and dashed curves are fits to the data according to Eqs.~\ref{T1Eqn1} and \ref{T1Eqn2}.}
\label{T1}
\end{figure}
Figure~\ref{T1} shows our measurements of the relaxation time for TEMPO-doped butanol at 0.56~T for temperatures
28--150~mK.  Included are data for butanol doped with a similar spin density of porphyrexide measured with the
Bonn frozen spin target at 0.42~T \cite{BradkteThesis}.  We have fit both sets of data according to a semi-empirical
relationship developed by de~Boer \cite{deBoer1974}
\begin{equation}
T^{-1}_{1p} = [A T_{1e} B^2 \cosh^2(\frac{h \nu_e}{2kT})]^{-1} + [a \frac{B^b}{T^c}]^{-1}
\label{T1Eqn1}
\end{equation}
where
\begin{equation}
AT_{1e} = 225[B^5 \coth(\frac{h \nu_e}{2kT}) + 6.75 \times 10^5 \exp(\frac{-1}{2T})]^{-1}
\label{T1Eqn2}
\end{equation}
In these equations $T_{1p}$ and $T_{1e}$ are the spin-lattice relaxation times (in hours) for the proton and electron
respectively.  The magnetic field strength (in kG) is $B$, and $\nu_e$ is the corresponding electron
Larmor frequency. $A$ is a constant, and $a$, $b$ and $c$ are fitting parameters.  Reasonable agreement between both
sets of data with the above equations could be found with $a=3.1\times10^{-4}$, $b=4.3$, and $c=2.5$.
\begin{figure*}
\begin{center}$
\begin{array}{cc}
\includegraphics[width=2.5in]{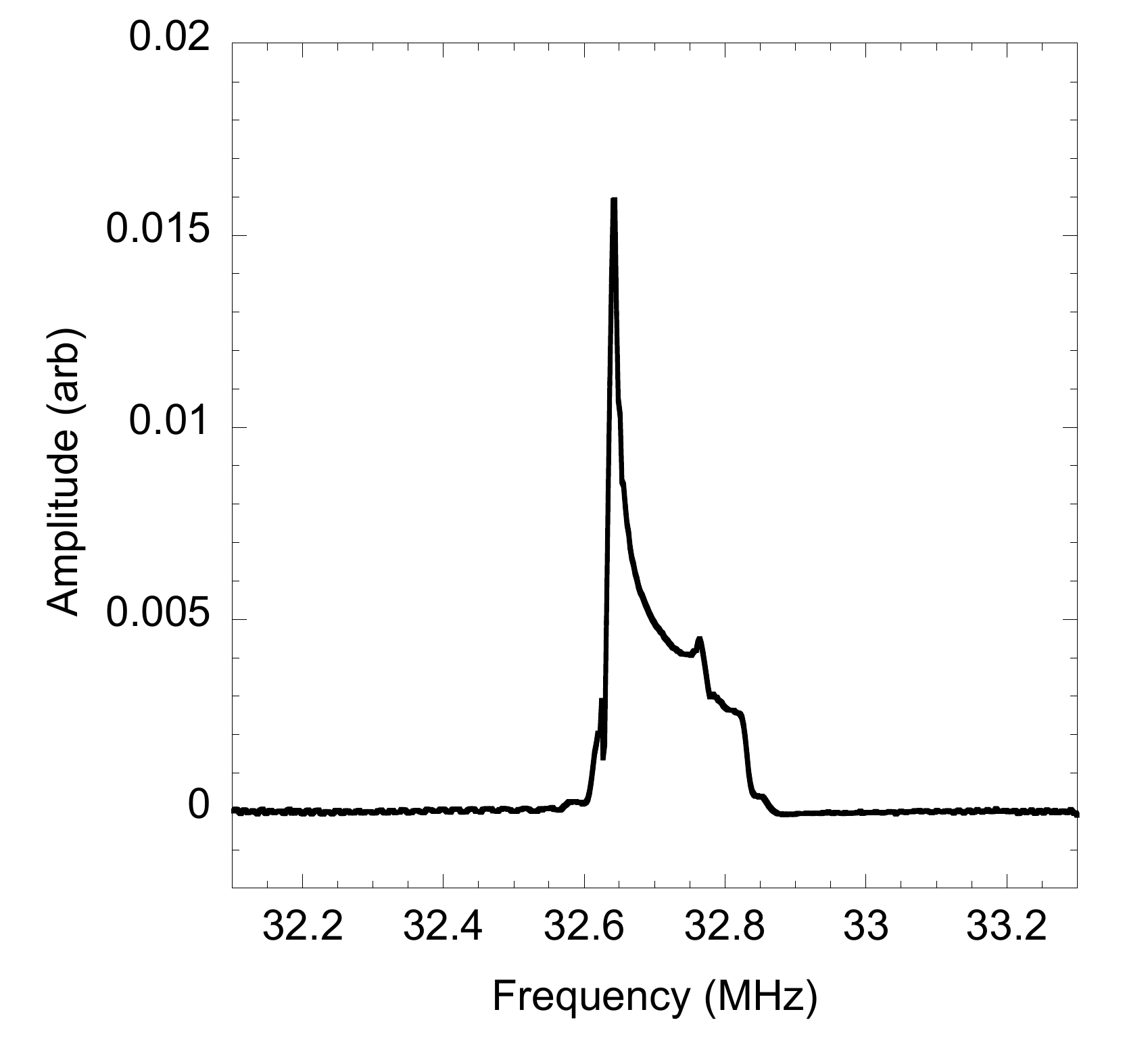} &
\includegraphics[width=2.5in]{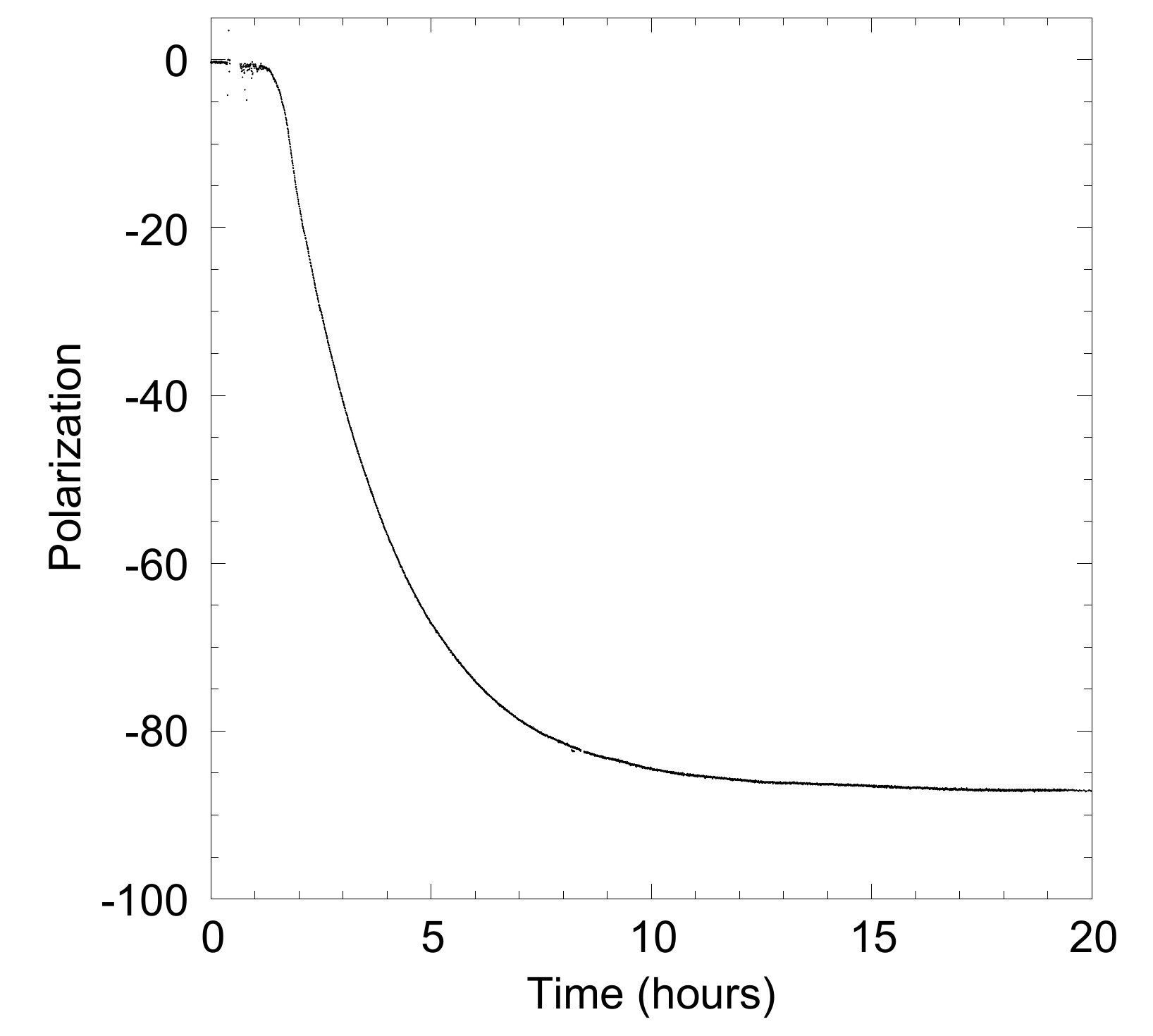}
\end{array}$
\end{center}
\caption{NMR signal and polarization growth curve of deuterated propanediol doped with OXO63.}
\label{TritylNMR}
\end{figure*}

Two separate NMR coils were used during g9a:  a single-loop coil 
connected to a Q-meter tuned to 212~MHz for measuring the
target polarization during DNP, and a two-loop coil and separate Q-meter at 24~MHz
for monitoring the polarization in frozen-spin mode.
The latter system was calibrated against the first each time the target was polarized, and the two
were again compared after approximately seven days of beam time, just before the target was re-polarized.
On average, the low frequency NMR system reported 4\% more polarization loss than
the high field system.  We attribute this discrepancy to nonuniform heating induced by the photon beam.
For geometrical reasons the two-loop coil was slightly more sensitive to material in the
downstream end of the target sample, where heating and depolarization
from forward-going charged particles was greatest.
During g9b the same two-loop coil was utilized for both the high- and low-frequency NMR
measurements, and no such discrepancy was observed.

\subsection{Polarization of Deuterated Propanediol}
\label{PolDProp}
At the conclusion of g9b, the butanol target was replaced with a sample of fully-deuterated propanediol, again
consisting of 1--2~mm frozen beads.  This sample was doped with the trityl
radical OX063\footnote{Oxford Scientific Instruments} with a spin concentration
of $1.5 \times 10^{19}$~cm$^{-3}$.  The newly synthesized trityl radicals have extremely narrow EPR
linewidths \cite{Heckmann06}, and deuteron polarizations as high as 80\% have been observed in both d-butanol
and d-propanediol \cite{Goertz04}.  To our knowledge no attempt has been
made to polarize trityl-doped samples at the operating conditions of FROST, 0.3~K and 5~T.  
The results of our test are shown in Figure~\ref{TritylNMR}.
A maximum deuteron polarization of -87$\pm 3$\% was obtained after approximately 16 hours of microwave
irradiation, while -80\% was obtained after about 7 hours.

It should be stressed that only one attempt was made to polarize a trityl-doped sample,
and we feel that even higher deuteron polarizations are possible.  It is likely that the spin concentration, which
was found to give the best results at 2.5~T \cite{Goertz04}, was not optimal for 5~T.  Further studies of
polarization with different spin densities will be performed in the future.

\section{Summary}
\label{Summary}
We have described a frozen spin polarized target constructed for use inside the CEBAF Large
Acceptance Spectrometer at Jefferson Lab.  The primary components of the target are a horizontal dilution
refrigerator with a high cooling power for dynamic polarization and a base temperature of about 25~mK,
and internal, superconducting magnets for maintaining the target polarization while in frozen spin mode.
Two such magnets have been used:  a 0.56~T solenoid for longitudinal polarization, and 
a 0.50~T saddle coil for transverse polarization.  The target has provided proton polarizations
greater than 90\%, and spin-lattice relaxation times up to 3400 hours have been observed with beam on target.
The target has been used with both linearly- and circularly-polarized photons to
perform the first ``complete'' photoproduction experiments at Jefferson Lab.
The high reliability of the target system, along with the long relaxation times,
resulted in an on-beam efficiency greater than 90\% during these experiments.
We have also demonstrated a deuteron polarization 
in this system of -87\% in trityl-doped d-propanediol, with the expectation of even
greater polarizations in the future.





\section{Acknowledgements}
The authors gratefully acknowledge the expert support 
provided by the technical and engineering 
staffs of the Jefferson Lab Target Group and Jefferson Lab 
Experimental Hall B during the design, construction and
operation of this target.

Authored by Jefferson Science Associates, LLC under 
U.S. DOE Contract No. DE-AC05-06OR23177. 
The U.S. Government retains a non-exclusive, paid-up, irrevocable, world-wide 
license to publish or reproduce this manuscript for U.S. Government purposes.

\bibliographystyle{model1a-num-names}
\bibliography{<your-bib-database>}


\end{document}